\documentclass[12pt]{article}
\usepackage{hyperref}
\usepackage{amsmath}
\usepackage{graphicx}
\usepackage{amssymb} 
 \usepackage{xcolor}
\definecolor{myblue}{rgb}{0.14,0.11,0.49}
\definecolor{myred}{rgb}{0.74,0.22,0.15}
\definecolor{mygreen}{rgb}{0.05,0.52,0.42}
\definecolor{myyellow}{rgb}{0.96,0.92,0.13}
\definecolor{myorange}{rgb}{1,0.61,0.36}
\definecolor{mypurple}{rgb}{0.71,0.02,1}
 
\newcommand{\Mat}[1]{{{\boldsymbol{#1}}}}
\newcommand{\abs}[1]{\left\vert#1\right\vert}
\def\be{\begin{equation}}
\def\ee{\end{equation}}
\def\bea{\begin{eqnarray}}
\def\eea{\end{eqnarray}}
\def\bi{\begin{itemize}}
\def\ei{\end{itemize}}
\def\noi{\noindent}
\def\dd{\mathrm{d}}
\def\iC{\mathrm{i}}

\date{}

\begin{document}
\title{Some remarks on quantum mechanics in a curved spacetime, especially for a Dirac particle}

\author{\small Appeared as: Int.  J. Theor. Phys. {\bf 54}, No. 7, 2218-2235 (2015), \href{http://link.springer.com/article/10.1007/s10773-014-2439-4}{DOI: 10.1007/s10773-014-2439-4}\\
\small\\
Mayeul Arminjon\\
\small\it Laboratory ``Soils, Solids, Structures, Risks'', 3SR,\\ \small\it CNRS and Grenoble-Alpes University,\\ \small\it BP 53, F-38041 Grenoble cedex 9, France.}

\maketitle


\begin{abstract} 
\noi Some precisions are given about the definition of the Hamiltonian operator H and its transformation properties, for a linear wave equation in a general spacetime. In the presence of time-dependent unitary gauge transformations, H as an operator depends on the gauge choice. The other observables of QM and their rates also become gauge-dependent unless a proper account for the gauge choice is done in their definition. We show the explicit effect of these non-uniqueness issues in the case of the Dirac equation in a general spacetime with the Schwinger gauge. We show also in detail why, the meaning of the energy in QM being inherited from classical Hamiltonian mechanics, the energy operator and its mean values ought to be well defined in a general spacetime.

\end{abstract} 
\section{Introduction and summary}

Quantum mechanics (QM) in a curved spacetime, thus in a gravitational field, is the domain that covers all currently available experiments about the interaction between gravity and the quantum. An important part of QM in a curved spacetime is QM of the covariant Dirac equation. The formulation of that equation involves the data of an arbitrary tetrad field on the spacetime, that is a field of orthonormal bases (e.g. \cite{BrillWheeler1957, ChapmanLeiter1976}). Two choices of the tetrad field lead to equivalent Dirac equations \cite{Fock1929}, at least locally. From this correct fact, it has been tacitly inferred, without proof, that the choice of the tetrad field has no physical effect whatsoever. This seems to remain the preferred opinion in the literature on QM of the Dirac equation in a gravitational field, in spite of recent work having proved that important objects such as the Hamiltonian operator $\mathrm{H}$, as well as its Hermitian part or energy operator, do depend on the tetrad field \cite{A43}. The dependence on the tetrad field includes that of the spectrum of the energy operator $\mathrm{E}$ \cite{A43}, that of the mean values of both $\mathrm{H}$ and $\mathrm{E}$ \cite{A50}, and it includes also the presence or absence of the spin-rotation coupling term \cite{Ryder2008,A49}. The aim of the present work is to discuss three important complementary points, two of which pertain to curved-spacetime QM in general, and which are strongly related together:\\

\noi {\bf 1.} In {\bf Section \ref{Hamiltonians}} we discuss the definition of the Hamiltonian $\mathrm{H}$ for a general wave equation in a general spacetime. We note that it naturally depends on the coordinate system, but merely through the corresponding reference frame --- the latter notion being defined as an equivalence class of coordinate systems. (We recall this definition and emphasize its strong link with the usual notion of a reference frame.) Then we show that, for any wave equation which transforms covariantly under a unitary ``gauge" transformation $\,\mathcal{U}\,$ depending on the time coordinate, $\mathrm{H}$, as an operator acting on the quantum-mechanical states, is not unique in a given reference frame. This non-uniqueness is there despite the fact that $\mathrm{H}$, seen as the generator of the time evolution of the wave function, transforms consistently: it actually {\it follows} from this consistency and the time-dependence of $\,\mathcal{U}$. We show that in the presence of such a gauge choice, the definition of any other quantum-mechanical operator $\,\mathcal{O}\,$ has to be reconsidered in order to lead to gauge-independent mean values. We show that, only when this is the case, the {\it time evolution} of any quantum-mechanical operator is gauge-independent also.\\

\noi {\bf 2.} Recently, two groups of authors have explicitly claimed \cite{GorbatenkoNeznamov2011, GorbatenkoNeznamov2013} or implicitly suggested \cite{OST2013} that the non-uniqueness issue of the curved-spacetime Dirac theory is absent if one restricts the choice of the tetrad field by using the ``Schwinger gauge". If this were true, that would not disprove the existence of the non-uniqueness problem. That would merely mean that another prescription for solving it would exist --- namely, using the Schwinger gauge. (Indeed, two different such prescriptions have been presented \cite{A47,A48}.) But the claim \cite{GorbatenkoNeznamov2011, GorbatenkoNeznamov2013} has been plainly refuted in Ref. \cite{A50}. Besides a general proof that the mean values of the Hermitian Hamiltonian provided by the construction \cite{GorbatenkoNeznamov2011,GorbatenkoNeznamov2011b} depend on the choice of the Schwinger tetrad, that work has shown explicitly this dependence in the case of an {\it inertial} reference frame in a {\it Minkowski} spacetime. However, a later paper \cite{OST2013} states this: ``Tetrads (coframes) are naturally defined up to a local Lorentz transformation, and one usually fixes this freedom by choosing a gauge. We discussed the choice of the tetrad gauge in [40] and have demonstrated that a physically preferable option is the Schwinger gauge." This might be taken to mean that in the Schwinger gauge there is no ambiguity any more, and that the ambiguity is thus solved in a ``physical" way. We show in {\bf Section \ref{OST}} that in fact the important quantum-mechanical objects of the work \cite{OST2013} do depend on the freedom that remains in the Schwinger gauge, which freedom is fully present in the approach \cite{OST2013}. Using the results of Sect. \ref{Hamiltonians}, we show indeed that not only the Hermitian Hamiltonians but also the {\it spin operators} of Ref. \cite{OST2013} depend on the choice of the Schwinger tetrad. This dependence obscures the physical meaning of the equations of that paper. In passing, we disprove the attempt made in Ref. \cite{GorbatenkoNeznamov2014} to refute our former explicit example \cite{A50}, and we show that this explicit example extends to the general situation envisaged in Ref. \cite{OST2013}. 
\\

\noi {\bf 3.} Regarding the non-uniqueness issue of the energy operator and its spectrum and mean values: some physicists wonder if in a non-stationary gravitational field the mean values of the energy operator should really have a physical meaning, given that then these mean values evolve with time and there is no stationary state in the strict physical sense. This objection has been partly answered, in advance, in Ref. \cite{A43}. It is related with a conception according to which every appearance of an energy should refer to a conserved quantity. This should indeed be true for the {\it total energy} (in a sense and in circumstances which will be outlined) but it does not apply to the energy of a {\it test particle} in a time-dependent external field, which is relevant here. This objection will be answered in {\bf Section \ref{Energy}} by recalling that in QM the meaning of the energy is inherited from classical Hamiltonian mechanics, and by depicting in some detail the main features of the classical Hamiltonian energy: in particular, it is in general {\it not} conserved, and it depends exactly on the reference frame as we define it. The same is true for the quantum-mechanical energy as it arises from the energy operator $\mathrm{E}$ --- when the latter is well defined. We contrast the dependence of $\mathrm{E}$ on the reference frame with its dependence on the tetrad field (for the Dirac equation): whereas the former is natural, just as is the dependence of the energy of a {\it classical} particle on the reference frame, we show that it is physically superfluous to add the data of a tetrad field to the data of a reference frame.

\section{Hamiltonians in the presence of a gauge choice}\label{Hamiltonians}

\paragraph{1. Hamiltonian in a general spacetime. Dependence on reference frame. ---}\label{Def H} How is the Hamiltonian operator precisely defined in a general spacetime $\mathrm{V}$? Consider a quantum wave equation for a wave function $\varphi $ with several components (i.e., living in some vector bundle with base $\mathrm{V}$). When a local coordinate system is given, and also a local frame field for the wave function $\varphi $, the wave function gets a local expression $\Psi \equiv (\Psi ^a)$, with $\Psi ^a$ the components of $\varphi $ in the given frame field, and the equation itself gets a local expression. Assume that the latter contains only first-order derivatives $\Psi ^a_{,0}\equiv \partial _0\Psi ^a$ with respect to the time coordinate $x^0$, in such a way that the corresponding matrix $M$, with $M\Psi_{,0} =(M^a_{\ \,b} \Psi ^b_{,0})$, is invertible. In that case, and only in that case, the wave equation can be uniquely rewritten as the Schr\"odinger equation (setting $\hbar=1$ in this section):
\be \label{Schrodinger-general}
\mathrm{i} \frac{\partial \Psi }{\partial t}= \mathrm{H}\Psi, 
\ee
where the Hamiltonian operator $\mathrm{H}$ is a linear differential operator that contains no derivative with respect to the coordinate time $t\equiv x^0/c$. From this definition, it follows that $\mathrm{H}$ {\it a priori} depends on the coordinate system.
\footnote{\
$\mathrm{H}$ depends also on the frame field, but this will be kept fixed in this paper; when the wave function $\varphi $ takes values directly in a $\mathbb{C}^m$ space, as is the case for the standard covariant Dirac equation and some alternative versions \cite{A46}, the canonical basis of $\mathbb{C}^m$ is a preferred frame field, which it makes little sense to change.
} 
When this is fixed, the operator $\mathrm{H}$ in general still depends on the value of $t$, because so do in general the coefficients of the wave equation. Moreover, Eq. (\ref{Schrodinger-general}) is valid only in the domain of definition of the coordinate system, say $\mathrm{U}$, which is an open subset of the spacetime $\mathrm{V}$. (We assume that the frame field on which $\varphi $ is expanded is defined above $\mathrm{U}$.)\\

The transformation properties of $\mathrm{H}$ on changing the coordinate system can be deduced from the Schr\"odinger equation (\ref{Schrodinger-general}), see Ref. \cite{A47} and references therein. After such a change, one gets a different operator $\mathrm{H}'$ which in general is not related simply to $\mathrm{H}$. However, $\mathrm{H}$ acts as a unique operator inside any set of charts (coordinate systems) which exchange by a purely spatial change: 
\be\label{purely-spatial-change}
x'^j=f^j((x^k))\ (j,k=1,2,3)\ \mathrm{(or} \ {\bf x}'={\bf f}({\bf x})),\qquad \mathrm{and}\ x'^0=x^0. 
\ee
E.g., when the wave function behaves as a scalar under a coordinate change:\\ $\Psi'((x'^\mu ))=\Psi((x^\nu ))$, we have after a change (\ref{purely-spatial-change}) \cite{A47}:
\be\label{H'=H}
(\mathrm{H}'\Psi')((x'^\mu ))=(\mathrm{H}\Psi)((x^\nu )). 
\ee

\paragraph{We specifically call ``reference frame" an equivalence class $\mathrm{F}$ of charts}\label{Def-Ref-Frame} which are defined on the same open subdomain $\mathrm{U}$ of the spacetime and which exchange two by two by a purely spatial transformation of coordinates (\ref{purely-spatial-change}) \cite{A44}. The data of a reference frame determines a congruence of world lines. Namely, in any chart $\chi $ that belongs to the class $\mathrm{F}$, each line $l_{\chi \,{\bf x}}$ of the congruence is the set of the events $X$ in $\mathrm{U}$ that have the same given spatial position ${\bf x}\equiv (x^j)$:
\be\label{lines of M}
l_{\chi \,{\bf x}}=\{X\in \mathrm{U};\quad \chi (X)=(x^0,{\bf x}) \mathrm{\ for\ some\ }x^0 \  (\mathrm{that\ depends\ on\ }X\in l_{\chi \,{\bf x}}) \}.
\ee
Each among the lines $l_{\chi \,{\bf x}}$ is invariant under a change of chart that is internal to a given reference frame, i.e., we have obviously $l_{\chi' \,{\bf f}({\bf x})}=l_{\chi \,{\bf x}}$ if $\chi $ and $\chi '$ exchange by (\ref{purely-spatial-change}). The data of a reference frame $\mathrm{F}$ amounts to the data of the domain $\mathrm{U}$ of the spacetime, the congruence of the world lines (\ref{lines of M}), {\it and} the data of a time coordinate map $X \mapsto x^0(X)$ defined for all events $X\in \mathrm{U}$. Call $\mathrm{F}$ ``admissible" if the metric component $g_{00}>0$ in $\mathrm{U}$ in one chart $\chi \in \mathrm{F}$ (hence also in any other chart $\chi' \in \mathrm{F}$). Then the world lines (\ref{lines of M}) of $\mathrm{F}$ are time-like. {\it This notion \cite{A44} is a natural extension of the notion of reference frame of classical mechanics.} Indeed, in classical mechanics, a reference frame is usually defined as the set of the trajectories of the points of a rigid solid (the rigidity being with respect to the invariant Euclidean metric of the Galilean spacetime), plus the implicitly present absolute time. The extension relies in the following: (i) no rigidity is asked for, so the reference frame is a deformable one in the most general way; (ii) a general spacetime is considered; (iii) the time is just a coordinate map $X \mapsto x^0(X)$ and has nothing absolute.\\

Hence, the dependence of the Hamiltonian operator $\mathrm{H}$ on the coordinate system expresses precisely its dependence on the reference frame as it is defined thus. The quantum-mechanical {\it states,} on which the operator $\mathrm{H}$ is acting, are functions $\Psi $ of the triplet of spatial coordinates, $ {\bf x} \equiv (x^j)\ (j=1,2,3)$, hence they are restricted to the spatial range $\Omega $ of the chart $\chi $. In order that $\Omega $ be independent of $t$, we may restrict the chart $\chi$ to a domain of the form: $\mathrm{U}'=\chi ^{-1}(\mathrm{I}\times \Omega)\subset  \mathrm{U} $, with $\mathrm{I}$ an open interval and $\Omega $ an open subset of $\mathbb{R}^3$.
\footnote{\label{SpaceManifold}\ 
In a more satisfying way, one can regard the states $\Psi $ as functions defined on the {\it space manifold} $\mathrm{M}$ associated \cite{A44} with the reference frame $\mathrm{F}$ fixed by the starting coordinate system. The elements (points) of $\mathrm{M}$ are the world lines of the form  (\ref{lines of M}). After a coordinate change of the form (\ref{purely-spatial-change}), which by definition leaves us in the same reference frame, the space manifold  $\mathrm{M}$ is left unchanged, and so is therefore the space of states. Also, $\mathrm{M}$ and hence the space of states do not depend on the time coordinate $x^0$.
} 
The fact that the Hamiltonian operator is defined only over the domain $\mathrm{U}$ of a chart has hardly any practical consequence, because in practice $\mathrm{U}$ can be taken such that $\chi (\mathrm{U})=\mathrm{I}\times \Omega $ with $\Omega $ large enough, so that the wave function $\Psi (t,{\bf x})$ can safely be assumed to vanish when ${\bf x}$ is at the boundary of $\Omega $, in a relevant time interval $\mathrm{I}$. In Ref. \cite{A49}, after Eq. (26), the example of a rotating frame is discussed. As usual, we adopt the same notation $\Psi $ for the state (a function of the spatial position ${\bf x} \in \Omega $) and for (the local expression of) the wave function, although strictly speaking the latter is defined on  $\mathrm{U}$ or equivalently on $\chi (\mathrm{U})$, i.e. it depends on the coordinate time $t\in \mathrm{I}$ and on  ${\bf x} \in \Omega $. 
\footnote{\
Multi-particle states are defined on a Cartesian product $\Omega ^n\equiv \Omega \times ...\times \Omega $. Then the wave function is a function defined on $\mathrm{I} \times \Omega^n $.
}

\paragraph{2. Effect of a gauge change on $\mathrm{H}$. ---} Until Section \ref{Energy} not included, we fix the \hyperref[Def-Ref-Frame]{reference frame}. In particular, the time coordinate map $X\mapsto x^0(X)=ct(X)$ is fixed. We allow that the linear differential operator which defines the wave equation involve the choice of some ``gauge" fields collectively denoted by $G$. (That choice was assumed done at \hyperref[Def H]{Point {\bf 1}}. For the covariant Dirac equation, $G$ would be the tetrad field.) We assume:  ({\bf i}) that, for any possible choice of the gauge fields $G$, there is a scalar product: $(\Psi ,\Phi )\mapsto (\Psi \mid \Phi )$, defined on a subset of the set of the states, which makes this subset a Hilbert space ${\sf H}$. [In practice, ${\sf H}$ is made of the states $\Psi $ for which $(\Psi \mid \Psi )$, defined as some space integral, is a well-defined finite number. The scalar product evolves in general with $t$, due to the time dependence of the coefficient fields in that space integral, but if that dependence is lower and upper bounded, ${\sf H}$ remains constant. This is not necessary in what follows, i.e., it might be the case that the space ${\sf H}$ itself evolve with the time $t$.] Thus, if $\widetilde{G}$ is another possible choice for the gauge fields, there is also a scalar product, which in general is different, say $(\Xi  ,\Omega  )\mapsto (\Xi \, \widetilde{\mid}\, \Omega  )$, and which defines another Hilbert space, say $\widetilde{{\sf H}}$. Moreover, we assume ({\bf ii}) that, with any such change of gauge fields, is uniquely associated at any time $t$ a {\it unitary transformation} $\mathcal{U}$ (evolving in general with $t$) between the respective Hilbert spaces ${\sf H}$ and $\widetilde{{\sf H}}$, $\Psi \mapsto \widetilde { \Psi } =\mathcal{U}\Psi$:
\be \label{tilde=isometry}
 \forall t\in \mathrm{I},\ \forall \Psi ,\Phi \in {\sf H},\quad  (\mathcal{U}\Psi \,\widetilde { \mid}\,\mathcal{U}\Phi )\equiv  (\widetilde { \Psi }  \ \widetilde { \mid} \ \widetilde { \Phi }  ) =(\Psi \mid \Phi ).
\ee
Lastly, we assume ({\bf iii}) that the wave equation changes covariantly under any possible change of gauge fields: $G \hookrightarrow \widetilde{G}$, thus under any admissible unitary transformation $\mathcal{U}$. Since the wave equation is equivalent (in the domain $\mathrm{U}$) to the Schr\"odinger equation (\ref{Schrodinger-general}), this means that, noting $ \widetilde{\mathrm{H}}$ the Hamiltonian corresponding with the new gauge fields $\widetilde{G}$, the starting Schr\"odinger equation (\ref{Schrodinger-general}) is equivalent, in the domain $\mathrm{U}$, to 
\be \label{Schrodinger-general-tilde}
\mathrm{i} \frac{\partial \Xi  }{\partial t}= \widetilde{\mathrm{H}}\Xi ,
\ee
where the wave function $\Xi $ in the new representation is deduced from the wave function $\Psi $ in the old representation by
\be\label{Xi = U Psi}
\Xi \equiv \mathcal{U}\Psi .
\ee 
These three assumptions, taken together, seem to be necessary in order that the possibility of a ``gauge choice"  in the wave equation may remain compatible with a consistent quantum mechanics for that equation. (Unfortunately, they are not sufficient for this.) They are valid for the (standard form of the) covariant Dirac equation \cite{A50}.\\

Substituting $\Psi=\mathcal{U}^{-1} \Xi$ into (\ref{Schrodinger-general}) and identifying with (\ref{Schrodinger-general-tilde}), we get easily:
\be\label{Htilde}
\widetilde{\mathrm{H}}=\mathcal{U} \mathrm{H}\mathcal{U}^{-1}-\mathrm{i}\,\mathcal{U}\left[\partial_t \left(\mathcal{U}^{-1}\right)\right]. 
\ee
This is indeed the well-known relation between two Hamiltonians exchanging by the most general ``operator gauge transformation" \cite{FoldyWouthuysen1950,Goldman1977}. 
Thus, when regarded as the generator of the time evolution of the wave function (the r.h.s. in the Schr\"odinger equation), the Hamiltonian transforms consistently, in the sense that the time evolution of $\Psi$ calculated with $\mathrm{H}$ and that of $\Xi $ calculated with $\widetilde{\mathrm{H}}$ are related by (\ref{Xi = U Psi}), at any time $t$.  On the other hand, when regarded as an operator acting on the states, the Hamiltonian operator, {\it and in fact any operator,} would transform consistently if, at any time $t$, all scalar products of the form $(\Psi \mid \mathrm{H} \Phi )$ were invariant under the unitary transformation $\mathcal{U}$, i.e., if one would have 
\be \label{H tilde equivt to H}
\forall \Psi ,\Phi \in \mathcal{D},\quad (\mathcal{U}\Psi\ \widetilde{\mid}\ \widetilde{\mathrm{H}} \, (\mathcal{U}\Phi) ) = (\Psi \mid \mathrm{H} \, \Phi ).
\ee
[Here, $\mathcal{D}$  is the domain (of definition) of the operator $\mathrm{H}$. The domain $\mathcal{D}$ is a linear subspace of the whole Hilbert space $\mathsf{H}$, and $\mathcal{D}$ should be dense in $\mathsf{H}$.] Note that this condition involves, in particular, the property that all mean values be invariant under the unitary transformation:
\be \label{H tilde equivt to H - mean}
\forall \Psi \in \mathcal{D},\quad \langle \widetilde{\mathrm{H}} \rangle \equiv  (\mathcal{U}\Psi\ \widetilde{\mid}\ \widetilde{\mathrm{H}} \, (\mathcal{U}\Psi) ) = (\Psi \mid \mathrm{H} \, \Psi )\equiv \langle \mathrm{H} \rangle.
\ee
{\it Independently of the nature of the operator} $\mathrm{H}$ (i.e., whether it is the Hamiltonian or not), it can be proved \cite{A50} that the conditions (\ref{H tilde equivt to H}) and (\ref{H tilde equivt to H - mean}) are actually equivalent, and that both are equivalent to the following relation between $\mathrm{H}$ and $\widetilde{\mathrm{H}}$:
\be\label{Htilde equivt to H - iff}
\widetilde{\mathrm{H}}=\mathcal{U} \mathrm{H}\mathcal{U}^{-1}. 
\ee
However, what we do have in general {\it for the Hamiltonian operator} is (\ref{Htilde}). The comparison between (\ref{Htilde}) and (\ref{Htilde equivt to H - iff}) proves that (\ref{Htilde equivt to H - iff}) is true iff $\partial_t \mathcal{U}=0$. Thus, in the generic situation that the unitary transformation $\mathcal{U}$ does depend on the coordinate time $t$, Eq. (\ref{H tilde equivt to H - mean}) is not true, i.e., it {\it cannot be the case} that the mean values of the Hamiltonians before and after the transformation be always equal for corresponding states. That the transformation (\ref{Htilde}) of the Hamiltonian is not in general compatible with the invariance of its mean values, has been noted a long time ago by Goldman \cite{Goldman1977}, without a proof, and in the simpler context of a Minkowski spacetime. He nevertheless provided a rather general example of a time-dependent transformation $\mathcal{U}$ for which this invariance [Eq. (\ref{H tilde equivt to H - mean}) here] indeed does not take place. 

\paragraph{3. Effect on other operators of QM. ---}\label{Other operators} The condition (\ref{Htilde equivt to H - iff}), which ensures that the mean values for corresponding states before and after the unitary transformation are equal, has the same form for any quantum-mechanical operator $\mathcal{O}$. To emphasize this, let us rewrite it so:
\be\label{Otilde equivt to O}
\widetilde{\mathcal{O}}=\mathcal{U} \mathcal{O}\mathcal{U}^{-1}. 
\ee
Thus, the usual definition of such an operator can remain valid only if it holds in the same way for the operator $\mathcal{U} \mathcal{O}\mathcal{U}^{-1}$, for whatever admissible unitary transformation. For instance, the usual definition of the momentum operator is:
\be\label{def-P_j}
P_j\equiv -\mathrm{i} \frac{\partial }{\partial x^j}\quad (j=1,2,3).
\ee
Consider the case that the admissible unitary transformations (associated with admissible changes of the gauge fields $G$) are given by local similarity transformations:
\be\label{psitilde=S^-1 psi} 
(\mathcal{U}\Psi)(X) \equiv S(X)^{-1}\Psi(X),
\ee
where $S(X)$ is some regular matrix depending smoothly on the spacetime position $X$. (This case includes the covariant Dirac theory \cite{BrillWheeler1957, ChapmanLeiter1976, A50}.) Clearly, with a unitary transformation $\mathcal{U}$ of the form (\ref{psitilde=S^-1 psi}), we do not have $\mathcal{U} P_j\mathcal{U}^{-1}=P_j$ in general. In other words, the mean values of the operator (\ref{def-P_j}) depend on the gauge choice, i.e. on the choice of the gauge fields $G$. Hence, if we maintain the definition (\ref{def-P_j}) for the momentum operator, then either: (i) this definition is valid only with particular gauge choices, that exchange two by two by similarity transformations which do not depend on the spatial coordinates; or: (ii) the mean values of the said momentum operator depend on the gauge choice.\\

The Hamiltonian operator also defines the time evolution of all quantum-mechanical operators: 
\be\label{def dO/dt}
\frac{\dd \mathcal{O}}{\dd t}\equiv  \frac{\partial  \mathcal{O}}{\partial  t}+\mathrm{i}  \left [\mathrm{H},\mathcal{O} \right ].
\ee
What happens in the presence of a gauge choice? Assume that, either by chance or by force, some quantum-mechanical observable $\mathcal{O}$ transforms as (\ref{Otilde equivt to O}). Consider first the case of a time-independent unitary transformation: $\partial _t\mathcal{U}=0$. Then, as we saw, the Hamiltonian operator transforms as (\ref{Htilde equivt to H - iff}), and we get immediately by applying (\ref{def dO/dt}):
\be\label{dOtilde/dt}
\frac{\dd \widetilde{\mathcal{O}}}{\dd t}\equiv  \frac{\partial  \widetilde{\mathcal{O}}}{\partial  t} + \mathrm{i}  \left [\widetilde{\mathrm{H}},\widetilde{\mathcal{O}} \right ]= \mathcal{U} \frac{\dd \mathcal{O}}{\dd t} \mathcal{U}^{-1}.
\ee
In the general case that $\partial _t\mathcal{U} \ne 0$, there are two new terms in $\,\mathrm{i} \, [\widetilde{\mathrm{H}},\widetilde{\mathcal{O}} ]$, due to the additional term $-\mathrm{i}\,\mathcal{U}\left[\partial_t \left(\mathcal{U}^{-1}\right)\right]$ in $\widetilde{\mathrm{H}}$, but those new terms are cancelled out by the two new terms in $\frac{\partial  \widetilde{\mathcal{O}}}{\partial  t}$, so Eq. (\ref{dOtilde/dt}) holds true. Thus the time derivative of the observable $\mathcal{O}$, as defined by (\ref{def dO/dt}), is gauge-independent. Obviously this is not the case, even when $\partial _t\mathcal{U}=0$, if $\mathcal{O}$ does not transform as (\ref{Otilde equivt to O}). In other words, an inappropriate account of the gauge choice in the definition of the observable transmits to its time derivative.

\section{Explicit consequences in a general metric}\label{OST}

In a recent paper \cite{OST2013}, Obukhov, Silenko \& Teryaev consider the covariant Dirac equation in an arbitrary coordinate system in a general spacetime. They derive quantum-mechanical equations and compare them with classical equations. They restrict the choice of the tetrad field by using the ``Schwinger gauge". Even with this restriction, a whole functional space of different tetrad fields still remains available. It has been shown \cite{A50,A47} that the Hermitian Hamiltonians deduced from two different choices of the ``Schwinger tetrad" by precisely the same non-unitary transformation of the wave function as the one used in Ref. \cite{OST2013} are in general physically inequivalent --- in the sense that their mean values for corresponding states are different. This occurs exactly for the general reason discussed in Section \ref{Hamiltonians}: two different choices of the Schwinger tetrad are in general related together by a time-dependent local Lorentz transformation, leading to a time-dependent unitary transformation between the wave functions before and after the change of tetrad. Let us thus examine in detail how that inequivalence issue affects the work \cite{OST2013}.

\paragraph{1. Non-uniqueness of a tetrad field in the Schwinger gauge. ---}\label{NU-Tetrad} With the parameterization of the spacetime metric used by the authors of Ref. \cite{OST2013} (hereafter OST for short), Eq. (2.1), the freedom in the choice of a Schwinger tetrad \cite{A47,GorbatenkoNeznamov2011b} appears fully, in the following way.
\footnote{\ 
All equation numbers of the form $(m.n)$ refer to the corresponding equations in Ref. \cite{OST2013}. To allow an easy reference, we mostly adopt the notations of that work, which differ from ours.
}
As noted by OST, ``the line element (2.1) is invariant under redefinitions
$W^{\widehat a}{}_b\longrightarrow L^{\widehat a}{}_{\widehat c}\,W^{\widehat c}{}_b$ using arbitrary local rotations $L^{\widehat a}{}_{\widehat c}(t,x)\in SO(3)$." [Clearly, here $x$ denotes the triplet $x\equiv (x^a)\ (a=1,2,3)$. Note that OST consider an arbitrary spacetime coordinate system, but do not envisage its change.] Under such a redefinition, the cotetrad field defined by Eq. (2.2): $\theta ^\alpha \equiv  e^\alpha _{\ \, i }\, dx^i \ (\alpha,i =0,...,3)$ changes to $\theta'^\alpha \equiv e'^\alpha_{\ \ i }\, dx^i$ [$(dx^i)$ being the basis of one-forms dual to the natural basis $(\partial _i)$ of the spacetime coordinate system $(x^i)$, where $x^0\equiv t$ according to OST's convention], with 
\be\label{theta'}
e'^{\widehat 0}_{\ \ i }\equiv e^{\widehat 0}_{\ \, i }\ \ (\theta'^{\widehat 0}\equiv \theta ^{\widehat 0}), \quad e'^{\widehat{a}}_{\ \ i} \equiv W'^{\widehat a}{}_b\left(\delta^b_i - cK^b\,\delta^{\,0}_i\right),
\ee
and where
\be\label{W'}
W'^{\widehat a}{}_b \equiv L^{\widehat a}{}_{\widehat c}\,W^{\widehat c}{}_b.
\ee
The tetrad field $u_\alpha \equiv e^i_{\ \, \alpha }\partial _ i$ defined by Eq. (2.3) changes to $u'_\alpha \equiv  e'^i_{\ \, \alpha } \partial _i$, with 
\be\label{u'}
e'^i_{\ \, \widehat 0 }\equiv e^i_{\ \, \widehat 0 }\ \ (u'_{\widehat 0} \equiv u_{\widehat 0}), \quad e'^i_{\ \, \widehat a }\equiv \delta ^i_b\, W'^b_{\ \ \widehat a },
\ee
where 
\be
W'^b_{\ \ \widehat a } \equiv W^b_{\ \, \widehat c }\,P^{\widehat c}_{\ \, \widehat a },
\ee
with $P=(P^{\widehat c}_{\ \, \widehat a }) \ (c, a =1,2,3)$ the inverse matrix of the matrix $(L^{\widehat a}{}_{\widehat c})$: $P=P(t,x)\in SO(3)$, hence $P^{\widehat c}_{\ \, \widehat a }=L^{\widehat a}{}_{\widehat c}$. Thus, the new tetrad field in the Schwinger gauge, Eq. (\ref{u'}), is deduced from the first one by a local Lorentz transformation $\Lambda =\Lambda (t,x)\in SO(1,3)$:
\be\label{Lambda}
u'_\beta =\Lambda ^\alpha _{\ \,\beta }\,u_\alpha , \qquad  \Lambda =\begin{pmatrix} 
1 & 0  & 0 & 0\\
0  &  &  & \\
0 &  & P &\\
0 &   &  &
\end{pmatrix}.
\ee 
Such a redefinition, envisaged by OST themselves, affects many relevant quantities. E.g. in (2.6), $\mathcal{F}^a_{\ \,b}$ becomes
\be\label{F'}
\mathcal{F}'^a_{\ \ b} = \mathcal{F}^a_{\ \,c} P^{\widehat c}_{\ \, \widehat b }.
\ee 
It affects also ${\mathcal Q}_{\widehat{a}\widehat{b}}$ in (2.11), ${\mathcal C}_{\widehat{a}\widehat{b}}{}^{\widehat{c}}$ in (2.12), hence also $\Gamma _{i\alpha \beta }$ in (2.9), (2.10) and (2.8), etc. Hence, {\it a priori,} every result in the ``quantum" part of Ref. \cite{OST2013} (sections II-III), as well as every comparison between quantum and classical results (section IV), may depend on the admissible choice of the field of the $3\times 3$ real matrix $W\equiv (W^{\widehat c}{}_b)$ --- that field being determined by the data of the spacetime metric only up to a spacetime dependent rotation matrix $(L^{\widehat a}{}_{\widehat c}(t,x))\in SO(3)$, Eq. (\ref{W'}).

\paragraph{2. Non-uniqueness of the Hermitian Hamiltonian operator (2.15). ---}\label{NU-H} As OST note, the non-unitary transformation (2.14) ``also appears in the framework of the pseudo-Hermitian quantum mechanics" as it is used in Ref. \cite{GorbatenkoNeznamov2011}. More precisely, with the usual relation $g^{\alpha\beta}\, e^i_{\ \,\alpha}\, e^j_{\ \,\beta} = g^{ij}$, the ``Schwinger gauge" condition $e^0_{\ \,\widehat a}=0$ gives 
\be\label{a^0_0}
g^{\widehat 0 \widehat 0}\,(e^0_{\ \,\widehat 0})^2 =g^{00},
\ee 
thus with OST's convention $x^0=t$: $e^0_{\ \,\widehat 0}=c \sqrt{g^{00}}=1/V$ [Eqs. (2.3) and (4.33)]. Instead, with $x^0=ct$, (\ref{a^0_0}) implies $\mid e^0_{\ \,\widehat 0}\mid =\sqrt{\abs{g^{00}}}$ independently of the signature \cite{A50,GorbatenkoNeznamov2011}. Therefore, the transformation (2.14):
\be\label{T=a^{-1}I}
\psi =\left(\sqrt{-g}\,e^0_{\ \,\widehat 0}\right)^\frac{1}{2}\Psi 
\ee
is exactly the one used by Gorbatenko \& Neznamov \cite{GorbatenkoNeznamov2011} to transform the Hamiltonian which they note $\widetilde{H}$,  got with some Schwinger tetrad, into the Hermitian Hamiltonian noted $\mathrm{H}_\eta$ by them, Eqs. (67) and (72) in Ref. \cite{GorbatenkoNeznamov2011}. It is also the particular case of the local similarity transformation $T$ in Eq. (18) of Ref. \cite{A50}, corresponding with $S={\bf 1}_4$ (one starts from a Schwinger tetrad). A crucial property of the transformation (\ref{T=a^{-1}I}), that it brings the Hilbert-space scalar product to the ``flat" form \cite{A50,GorbatenkoNeznamov2011}:
\footnote{\
In Eq. (\ref{scalar product})$_1$, $\gamma ^{\widehat 0}$ is the ``$\alpha =0$" constant Dirac matrix (assumed to be ``hermitizing" as is standard), which is noted $\gamma ^{\natural 0}$ in Ref. \cite{A50}. Whereas, $\gamma ^0$ is the ``$i=0$" matrix of the {\it field of Dirac matrices in the curved spacetime,}  $\gamma ^i\equiv e^i_{\ \,\alpha }\,\gamma ^\alpha $ ($\gamma ^\mu \equiv a^\mu _{\ \,\alpha }\,\gamma^{\natural  \alpha} $ in the notation of Refs. \cite{A47,A50}). Whether one mentions it or not, the field $\gamma ^i$ can be defined as soon as one has a tetrad field and a set of constant Dirac matrices $\gamma ^\alpha $ valid for the Minkowski metric, and it depends on both. In fact, the field $\gamma ^i$, or at least the field of the $\gamma ^0$ matrix, is needed to define the scalar product --- as shown precisely by Eq. (\ref{scalar product})$_1$. The field $\gamma ^i$ is also there in the Dirac equation, though not explicitly with the form (2.7) used by OST: (2.7) rewrites using (2.8)$_1$ in the slightly more standard form  \be\label{Dirac-standard} (\mathrm{i}\hbar \gamma ^i D_i-mc)\Psi =0. \ee
}
\be \label{scalar product}
(\Psi  \mid \Phi  ) \equiv \int_\Omega \Psi^\dagger \sqrt{-g}\,\gamma ^{\widehat 0} \gamma ^0\, \Phi\ \dd^ 3{\bf x} \longrightarrow (\psi \, \widetilde{\mid} \,\phi  )= \int_\Omega \psi^\dagger \phi\ \dd^ 3{\bf x} ,
\ee
is ensured by Eq. (\ref{a^0_0}) above, due to Eq. (19) in Ref. \cite{A50}. (Here  $\Omega $ is the spatial range of the coordinate system, see Sect. \ref{Hamiltonians}.) Recall that the scalar product has to be specified before one can state that some operator is Hermitian. In this section, {\it Hermitian} operators are stated to be so with respect to the ``flat" product (\ref{scalar product})$_2$.\\

Thus, starting from one tetrad field $(u_\alpha )$ in the Schwinger gauge (2.2)--(2.3) and getting then, in general, a non-Hermitian Hamiltonian [with respect to the product (\ref{scalar product})$_1$], the Hermitian Hamiltonian $\mathcal{H}$ obtained by OST using the transformation (2.14) [Eq. (\ref{T=a^{-1}I}) here] is just the one denoted $\mathrm{H}_\eta$ in Refs. \cite{A50, GorbatenkoNeznamov2011}, with here $\eta = (\sqrt{-g}\,e^0_{\ \,\widehat 0})^\frac{1}{2} {\bf 1}_4$. Now, consider another tetrad field $(u'_\alpha )$ in the Schwinger gauge (2.2)--(2.3). It is thus related with the first one $(u_\alpha )$ by the local Lorentz transformation $\Lambda $, Eq. (\ref{Lambda}): essentially, $\Lambda $ is the arbitrary rotation field $P(t,x)=(L^{\widehat a}{}_{\widehat c})^{-1}\in SO(3)$. Define $S'$, the local similarity transformation got by ``lifting" the local Lorentz transformation $\Lambda $ to the spin group. At this point, we could repeat {\it verbatim} what is written in Ref. \cite{A50} after the first sentence following Eq. (19). Thus, let $'\mathcal{H}\equiv \mathrm{H}_{\eta'}$ be the Hermitian Hamiltonian obtained by OST using the transformation (\ref{T=a^{-1}I}), but starting from the Schwinger tetrad field $(u'_\alpha )$ instead of the other one $(u_\alpha )$. (The notation $\mathcal{H}'$ designates something else in Ref. \cite{OST2013}.) The change from the Hamiltonian $\mathcal{H}=\mathrm{H}_\eta$ to the Hamiltonian $'\mathcal{H}=\mathrm{H}_{\eta'}$ is through the local similarity transformation $U=S'S^{-1}$, Eq. (20) of Ref. \cite{A50}. (Here, $U=S'$, because we have $S={\bf 1}_4$ as noted after Eq. (\ref{T=a^{-1}I}) above.) This implies that the Schr\"odinger equation $\mathrm{i}\hbar\frac{\partial \psi} {\partial t} = {\mathcal H}\psi$ is equivalent to  $\mathrm{i}\hbar\frac{\partial \psi'} {\partial t}=\, '\mathcal{H}\,\psi'$, with $\psi'=U^{-1}\psi$, as with Eqs. (\ref{Schrodinger-general}) and (\ref{Schrodinger-general-tilde}) above: indeed, the covariant Dirac equation is covariant under local similarity transformations that belong to the spin group. Moreover, the similarity matrix $U$ is a unitary matrix. \{This property of the gauge transformations internal to the Schwinger gauge, derived in \cite{A50} from the invariance of the scalar product (\ref{scalar product})$_2$ under $U$, can be seen also from the fact that the Lorentz transformation (\ref{Lambda}) is a rotation.\} Hence the transformation
\be\label{psi'}
\psi \mapsto \psi '=U^{-1}\psi
\ee
is a unitary transformation internal to the Hilbert space $\mathsf{H}$. Here, in view of (\ref{scalar product})$_2$, $\mathsf{H}$ is the space of the square-integrable four-components complex functions of the spatial coordinates:
\be\label{H flat Dirac}
\psi \in \mathsf{H} \ \mathrm{iff}\ \psi : \Omega \rightarrow  \mathbb{C}^4 \quad \mathrm{is\ such\ that}\ (\psi \, \widetilde{\mid}\, \psi  )= \int_\Omega  \psi^\dagger \psi\ \dd^ 3{\bf x} <\infty.
\ee
{\it And\ } $'\mathcal{H}$ is physically inequivalent to $\mathcal{H}$, unless $\partial _t U=0$, i.e., unless the arbitrary rotation field $P$ is chosen independent of $t$. Indeed, we are in a particular case of the situation leading to Eq. (\ref{Htilde}), which writes here \cite{A43,A50,GorbatenkoNeznamov2011}:
\be\label{H' vs H}
'\mathcal{H}=U^{-1} \mathcal{H} U - \mathrm{i}\hbar U^{-1} \partial _t U,
\ee
with $U^{-1}=U^\dagger $, $U$ being a unitary matrix. 
\footnote{\ 
It is noted in Refs. \cite{A50, GorbatenkoNeznamov2011b} that $\mathcal{H}$ [respectively $'\mathcal{H}$] is also equal to the {\it energy operator} (the Hermitian part of the Hamiltonian) corresponding to the Schwinger tetrad $(u_\alpha )$ [respectively $(u'_\alpha )$]. Accounting for Eq. (\ref{a^0_0}) and for the fact that here $U$ is unitary, the same relation (\ref{H' vs H}) is got by using the general relationship \cite{A43} between two energy operators related by an admissible local similarity transformation. 
} 

\paragraph{3. Physical inequivalence of $\mathcal{H}$ and $ '\mathcal{H}$. ---}\label{H vs H'} As we saw in Section \ref{Hamiltonians}, and contrary to what is stated by Gorbatenko \& Neznamov \cite{GorbatenkoNeznamov2013,GorbatenkoNeznamov2014}, this inequivalence (in the sense of the mean values) is what is expressed by Eq. (\ref{H' vs H}), unless $\partial _t U=0$. Recall that the mean value $\langle \mathcal{H} \rangle $ depends on the state $\psi $, belonging to the domain $\mathcal{D}$ of the operator $\mathcal{H}$. Here, in view of (\ref{scalar product})$_2$: 
\be\label{Def mean value}
\langle \mathcal{H} \rangle =\langle \mathcal{H} \rangle _\psi \equiv (\psi \, \widetilde{\mid}\, \mathcal{H}\psi  )= \int_\Omega \psi^\dagger (\mathcal{H}\psi)\ \dd^ 3{\bf x} \qquad \mathrm{when}\ \psi \in \mathcal{D}.
\ee
The precise definition of $\mathcal{D}$ should ensure that the integral above makes sense, for any $\psi \in \mathcal{D}$. The mean value $\langle '\mathcal{H} \rangle$ for the corresponding state $\psi '$ after the transformation (\ref{psi'}) is given by the same Eq. (\ref{Def mean value}), with primes. It has been proved in Ref. \cite{A50} that, if the similarity matrix $U(t,x)$ in Eq. (\ref{H' vs H}) depends indeed on $t$, then not only the mean values $\langle \mathcal{H} \rangle$ and $\langle '\mathcal{H} \rangle$ are in general different, but in addition the difference $\langle '\mathcal{H} \rangle - \langle \mathcal{H} \rangle$ depends on the state $\psi \in \mathcal{D}$. Moreover, the difference $\langle '\mathcal{H} \rangle - \langle \mathcal{H} \rangle$ can be calculated explicitly when the tetrad $(u_\alpha(t,x))$ is deduced from $(u'_\alpha (t,x))$ by the rotation of angle $\omega t$ around the vector $u_3(t,x)=u'_3(t,x)$. In that case, one has the explicit expression 
\be\label{U=S(Lambda)}
U(t)=e^{\omega tN},\qquad N\equiv (\alpha^1\alpha^2)/2\quad  (N^\dagger=-N).
\ee
Here, $\alpha^a \equiv  \gamma^{\widehat 0} \gamma^a$ ($a= 1,2,3$), in the notation of \cite{OST2013}, that is $\alpha '^j \equiv \gamma '^0\gamma '^j\ $ in the notation of Ref. \cite{GorbatenkoNeznamov2013}. Equation (\ref{U=S(Lambda)}) \cite{GorbatenkoNeznamov2013} applies to this more general situation as well. Indeed, the spin transformation $U$ (noted $R$ in Ref. \cite{GorbatenkoNeznamov2013}) that lifts the local Lorentz transformation (\ref{Lambda}) with $P^T$ the rotation of angle $\omega t$ around $u_3$ is independent of whether or not $u_3$ and $\partial _3$ coincide (as was the case in Ref. \cite{GorbatenkoNeznamov2013}). We get from (\ref{U=S(Lambda)})$_2$: $N=\frac{\iC }{2}\Sigma ^3\equiv \frac{\iC }{2}\mathrm{diag}(1,-1,1,-1)$ \cite{A50}, whence by (\ref{H' vs H}), (\ref{Def mean value}) and (\ref{U=S(Lambda)})$_1$ \cite{A50}:
\be\label{delta <H>}
\langle '\mathcal{H} \rangle - \langle \mathcal{H} \rangle =\frac{\hbar \omega}{2}\langle \Sigma ^3 \rangle
=\frac{\hbar \omega}{2}\int_\Omega \left(\abs{\psi^0}^2 +\abs{\psi^2}^2-\abs{\psi^1}^2-\abs{\psi^3}^2\right)\,\dd^3{\bf x}.
\ee
This even holds true in the presence of an electromagnetic field \cite{A49}. This equation applies to {\it any} possible state $\psi \in \mathcal{D}$. It implies that, for the states $\psi\in \mathcal{D}$ such that $ \langle \Sigma ^3 \rangle \ne 0$, i.e. the integral in (\ref{delta <H>}) does not vanish, then definitely $\langle '\mathcal{H} \rangle \ne \langle \mathcal{H} \rangle$. The difference, $\delta =\frac{\hbar \omega }{2} \langle \Sigma ^3 \rangle$, depends on the state $\psi $. For a normed state: $(\psi \widetilde{\mid }\psi )=1$, $\delta $ can take any value between $\frac{\hbar \omega }{2}$ and $-\frac{\hbar \omega }{2}$.  Note that $\omega $ is the rotation rate of the tetrad $(u_\alpha)$ with respect to $(u'_\alpha )$ and can be made {\it arbitrarily large}. \\

The foregoing discussion extends to a general metric the explicit counterexample given in Ref. \cite{A50} as part of our answer to the arguments of Ref. \cite{GorbatenkoNeznamov2013}. However, in Ref. \cite{GorbatenkoNeznamov2014}, which has otherwise nearly the same content as Ref. \cite{GorbatenkoNeznamov2013}, it is stated after Eq. (34) that ``when averaging the physical quantities for the spin particles, one should also perform averaging over the spin states with an appropriate change of the normalizing condition." This is misleading. Recall that here we consider states which are appropriate for the Dirac equation: they belong to $\mathcal{D}$, that is a dense subspace of the Hilbert space ${\sf H}$ of four-components functions given by (\ref{H flat Dirac}). Thus, the ``spin attribute" is included in the data of a state $\psi \in \mathcal{D}$, although a state $\psi \in \mathcal{D}$ has in general not a well-defined value for the observable ``helicity", that defines the ``spin state". This means that the mean value $\langle \mathcal{H} \rangle _\psi$ for a state $\psi \in \mathcal{D}$ does include the appropriate ``averaging over the spin states", namely over those that (in general) are mixed in the state $\psi $. Of course there are states for which $ \langle \Sigma ^3 \rangle=0$; e.g., to go in the direction of the calculation in Eq. (35) of Ref. \cite{GorbatenkoNeznamov2014}: when the metric is the Minkowski metric of a flat spacetime, the coordinates $(x^i)$ being Cartesian and $u_3$ being parallel to $Ox^3$, take an average state $\psi _\mathrm{av}\equiv (\psi _1+\psi _2)/\surd 2$, with $\psi _j\in  \mathcal{D}$ of the form $\varphi(x)A_j\ (j=1,2)$ where $\varphi(x)$ is a square-integrable scalar function and $A_j\in \mathbb{R}^4$ are the amplitude vectors of two plane wave solutions of the free Dirac equation, {\it with momentum parallel to $Ox^3$ i.e. to the rotation axis of the tetrad $(u_\alpha)$ with respect to $(u'_\alpha )$,} and with opposite helicities $\pm \frac{1}{2}$ (see Ref. \cite{Schulten2000}). (In contrast with the said calculation in Ref. \cite{GorbatenkoNeznamov2014}, such an average state has to be defined {\it before} the mean value is calculated, for the mean value is not linear.) In that particular metric as well as in a general one, the states for which $ \langle \Sigma ^3 \rangle=0$ are a very small subset of all physical states $\psi \in \mathcal{D}$. Moreover, by choosing another rotation field $P$, one may easily show that, also for the states having $ \langle \Sigma ^3 \rangle=0$, the energy mean values are indeterminate.

\paragraph{4. Inequivalence of $\mathcal{H}$ and $\mathcal{H}_\mathrm{FW}$ in the time-dependent case. ---}\label{FW} It has just been proved that, when the rotation field $P$ in Eq. (\ref{Lambda}) depends on $t$, the two Hermitian Hamiltonians (2.15) $\mathcal{H}$ and $'\mathcal{H}$ are inequivalent in the sense that their mean values are different. This inequivalence would transmit automatically to that of the corresponding Foldy-Wouthuysen (FW) Hamiltonians (3.11), say $\mathcal{H}_\mathrm{FW}$ and $'\mathcal{H}_\mathrm{FW}$ --- if the FW transformation (3.2) would lead to an equivalent Hamiltonian to the starting one, i.e., if $\mathcal{H}$ were equivalent to $\mathcal{H}_\mathrm{FW}$, and $'\mathcal{H}$ to $'\mathcal{H}_\mathrm{FW}$. However, the issue of inequivalence enters the scene an {\it additional time} here, because the unitary transformation $\mathcal{U}$ (3.2) has just the form (\ref{Htilde}), which, for the case $\partial _t \mathcal{U} \ne 0$, implies that the Hamiltonians before and after transformation do not have the same mean values. This issue may have been noted by Eriksen \cite{Eriksen1958}, perhaps even by Foldy \& Wouthuysen themselves \cite{FoldyWouthuysen1950,Goldman1977}. Eriksen \cite{Eriksen1958} limited the application of the FW transformation to ``non-explicitly time-dependent" transformations, i.e. indeed $\partial _t \mathcal{U} = 0$ in Eq. (\ref{Htilde}). That issue has been discussed in detail by Goldman \cite{Goldman1977}, for the FW transformation applied to the Dirac equation in a flat spacetime in Cartesian coordinates in the presence of an electromagnetic field. He noted explicitly that, in the time-dependent case, the transformed Hamiltonian [here $'\mathcal{H}$ in (\ref{H' vs H})] ``is unphysical (in the sense of EEV's)" (energy expectation values), given that (in his case) ``the EEV's of the original $\mathcal{H}$ [are supposed to] have physical meaning". \\

The FW transformation $\mathcal{U}$ (3.3) used in effect by OST has the form 
\be\label{U_FW-OST}
\mathcal{U}=\mathcal{N}.(\mathcal{N}^2)^{-1/2}\beta 
\ee
with $\mathcal{N}=\mathcal{N}(t)\equiv \beta \epsilon +\beta \mathcal{M}-\mathcal{O}$ a time-dependent operator (implicitly assumed to be Hermitian and have an inverse), where ${\cal H}=\beta {\cal M}+{\cal E}+{\cal O}$ is a decomposition, stated by OST, of the Hermitian Hamiltonian $\mathcal{H}$, Eq. (3.1);
\footnote{\
Any operator acting on four-component functions decomposes uniquely into ``even" and ``odd" parts, i.e. commuting and anticommuting with $\beta \equiv \gamma ^{\widehat 0}\equiv \mathrm{diag}(1,1,-1,-1)$ \cite{FoldyWouthuysen1950,Eriksen1958}. The decomposition (3.1) of the given operator ${\cal H}$ is hence unique for {\it any} even operator $\cal{M}$ which is also {\it given}. In the present case, $\cal{M}$ appears to be given by Eq. (3.8). Then one can easily express ${\cal E}$ and ${\cal O}$ in terms of the Hermitian operators ${\cal H}$, $\beta $ and ${\cal M}$, and check that $\mathcal{N}$ is indeed Hermitian.
}
and $\epsilon \equiv \sqrt{{\cal M}^2+{\cal O}^2}$. Assume for simplicity that there is a subdomain $\mathcal{D}_\mathrm{e}(t) \subset \mathcal{D}$ to which the restriction of $\mathcal{N}(t)$ has a  decomposition in eigenspaces:  $\mathcal{D}_\mathrm{e}=\oplus _j \mathrm{E}_j$ and $\mathcal{N}_{\mid \mathcal{D}_\mathrm{e}}=\Sigma _j \lambda _j\, \mathrm{Pr}_{\mathrm{E}_j}$ with $\lambda _j(t)$ nonzero reals and $\mathrm{Pr}_{\mathrm{E}_j}$ the orthogonal projection on the eigenspace $\mathrm{E}_j(t)$. Then, if $\psi \in \mathrm{E}_j$, we have $\mathcal{N}.(\mathcal{N}^2)^{-1/2}\psi =\varepsilon _j\,\psi $ with $\varepsilon _j\equiv \mathrm{sgn}(\lambda _j)=\pm1$. Hence, defining $\mathcal{D}_\mathrm{e}^+=\oplus _{\varepsilon _j=+1} \mathrm{E}_j$ and the like for $\mathcal{D}_\mathrm{e}^-$, we have $\mathcal{D}_\mathrm{e}=\mathcal{D}_\mathrm{e}^+\oplus \mathcal{D}_\mathrm{e}^-$ and $[\mathcal{N}.(\mathcal{N}^2)^{-1/2}]_{\mid \mathcal{D}_\mathrm{e}}=\mathrm{Pr}_{\mathcal{D}_e^+} -\mathrm{Pr}_{\mathcal{D}_e^-}$, whence $\mathcal{U}_{\mid \beta ^{-1}(\mathcal{D}_\mathrm{e}^\pm)}=\pm \beta $ and 
\be\label{U Proj}
\mathcal{U}_{\mid \beta ^{-1}(\mathcal{D}_\mathrm{e})}=\left(\mathrm{Pr}_{\mathcal{D}_e^+ }-\mathrm{Pr}_{\mathcal{D}_e^-} \right)\beta ,
\ee
with in fact $\beta ^{-1}=\beta $. Now, clearly the eigenspaces $\mathrm{E}_j$ of the operator $\mathcal{N}(t)$ evolve with time in a general metric, due to the time-dependence of the metric and that of the tetrad, hence $\mathcal{D}_\mathrm{e}^+$ and $\mathcal{D}_\mathrm{e}^-$ also evolve. So Eq. (\ref{U Proj}) confirms that $\mathcal{U}$ is time-dependent. Thus in the general case the FW transformation (3.3) depends on time, so that the FW Hamiltonian $\mathcal{H}_\mathrm{FW}$ (3.11) does not have the same mean values as the starting Hermitian Hamiltonian $\mathcal{H}$ (2.15). This is in addition to the inequivalence of two Hermitian Hamiltonians $\mathcal{H}$ and $'\mathcal{H}$ (2.15) got from two choices of the Schwinger tetrad, proved at Points  \hyperref[NU-H]{({\bf 2})} and \hyperref[H vs H']{({\bf 3})}, and which applies even in the case of a Minkowski spacetime in Cartesian coordinates.

\paragraph{5. Non-uniqueness of the spin operators. ---}\label{Spin operators} To derive the equation of motion of spin, OST consider the polarization operator $\Mat{ \Pi}\equiv \beta \Mat{ \Sigma }$, with $\Mat{ \Sigma }\equiv (\Sigma ^a)$, where $\Sigma^1 \equiv \iC\gamma^{\hat 2}\gamma^{\hat 3}, \Sigma^2 \equiv \iC\gamma^{\hat 3}\gamma^{\hat 1}, \Sigma^3 \equiv \iC\gamma^{\hat 1}\gamma^{\hat 2}$. This definition is given independently of the choice of the Schwinger tetrad field. After a change of the latter, we thus have 
\be\label{Pi'}
\Mat{ \Pi}' \equiv \beta \Mat{ \Sigma } =\Mat{ \Pi}.
\ee
This operator is used in the quantum equation of motion of spin (3.15):
\be
\frac{\dd \Mat{ \Pi}}{\dd t}=\frac{\iC}{\hbar}[{\mathcal H}_\mathrm{FW},\Mat{ \Pi}].
\ee
Thus, this equation uses the FW representation based on the unitary transformation (3.2) of the Hilbert space ${\sf H}$ defined in Eq. (\ref{H flat Dirac}). Here we denote the transformation (3.2) as $\psi _\mathrm{FW}=\mathcal{U}\psi $, with $\mathcal{U} $ in script font. A change of the Schwinger tetrad field introduces first another unitary transformation, denoted by $U$, Eq. (\ref{psi'}), leading to the new representation $\psi '$. In the $\psi '$ representation, we get a different FW transformation, say $\psi' _\mathrm{FW}=\mathcal{U}'\psi' $, from the different Hermitian Hamiltonian operator $'\mathcal{H}$ related with $\mathcal{H}$ by Eq. (\ref{H' vs H}). Hence, the two FW representations, arising from two different choices of the Schwinger tetrad, are related together by the following unitary transformation of ${\sf H}$:
\be
\mathcal{V}=\mathcal{U}'U^{-1}\mathcal{U}^{-1},\quad \psi' _\mathrm{FW}=\mathcal{V}\psi _\mathrm{FW}.
\ee
Since $\Mat{ \Pi}' =\Mat{ \Pi}$ in view of (\ref{Pi'}), the operator $\Mat{ \Pi}'$ has clearly no reason to verify the condition (\ref{Otilde equivt to O}) with this complicated unitary transformation $\mathcal{V}$, i.e., we have in general
\be
\Mat{ \Pi}' \ne \mathcal{V} \Mat{ \Pi} \mathcal{V}^{-1}.
\ee 
This means that, as a quantum-mechanical observable characterized by scalar products, in particular mean values, the operator $\Mat{ \Pi}$ depends on the gauge choice, that is, on the choice of the Schwinger tetrad. Moreover, as we saw at \hyperref[Other operators]{Point {\bf 3}} in Sect. \ref{Hamiltonians}, this gauge dependence transmits to its time derivative, thus to the spin operator $\Mat{\Theta }\equiv \frac{\dd \Mat{ \Pi}}{\dd t}$ of the equation of motion of spin (3.15), which uses Eq. (\ref{def dO/dt}) above. I.e., the scalar products $(\psi _\mathrm{FW}\,\widetilde{\mid }\,\Mat{\Theta }\phi _\mathrm{FW})$, and in particular the mean values $\langle \Mat{\Theta } \rangle= (\psi _\mathrm{FW}\,\widetilde{\mid }\,\Mat{\Theta }\psi _\mathrm{FW})$, depend on the choice of the Schwinger tetrad.\\

This is consistent with the explicit dependence of $\mathcal{F}^a_{\ \,b}$ on the choice of the Schwinger tetrad field, noted in Eq. (\ref{F'}). Thus, the three-vector operators of the angular velocity of spin precessing $\Mat{\Omega}_{(1)}$ and $\Mat{\Omega}_{(2)}$, Eqs. (3.16) and (3.17), depend {\it a priori} also on that choice. Also, {\it a priori,} this dependence is expected to survive in the semi-classical limit (3.19)--(3.20), unless the authors prove otherwise. However, we note that the operator $p_a\equiv  -i\hbar \frac{\partial }{\partial x^a}$, as well as the c-number $p_a$ given by (4.30), are independent of the choice. Therefore, using $PP^T={\bf 1}_3$, one finds that $\epsilon '$ in Eq. (3.21) is invariant under the change of the Schwinger tetrad involving the substitution (\ref{F'}), for
\be\label{F=F'}
\delta^{cd}{\mathcal F}'^a{}_c\,{\mathcal F}'^b{}_d
\,p_a\,p_b = \delta^{cd}{\mathcal F}^a{}_e\,P^e_{\ \,c}\,{\mathcal F}^b{}_f\,P^f_{\ \,d}\,p_a\,p_b = \delta^{ef}{\mathcal F}^a{}_e\,{\mathcal F}^b{}_f\,p_a\,p_b.
\ee
This applies whether in (3.21) one considers $p_a$ and $\epsilon '$ as operators or as c-numbers. It follows, using again (\ref{F=F'}), that the semi-classical velocity operator $\frac{dx^a}{dt}$ as given by the last member of Eq. (3.23) is invariant under the change of the Schwinger tetrad. The semi-classical velocity operator in the Schwinger frame (3.24) is covariant under the change of the Schwinger tetrad: $v'_a=P^b_{\ \,a}\, v_b$, if $\epsilon '$ is regarded as a c-number in (3.24).

\paragraph{6. Uniqueness of the classical limit. ---} Classical spinning particles are discussed in two independent parts of  Ref. \cite{OST2013}: Sects. IV A and IV B. In Sect. IV A, the degree to which ``the classical equation of the spin motion (4.22) agrees with the quantum equation (3.15) and with the semiclassical one (3.18)" does not appear very clearly: these are three rather complex expressions which do not coincide, and as noted above the quantum equation (3.15) is ambiguous. In Sect. IV B, we note that the classical spin rate $\Mat{\Omega }$ (4.36), as well as [using (\ref{F=F'}) with $p_a \rightarrow \pi_a$] the classical Hamiltonian (4.38), are independent of the choice of the Schwinger tetrad. In their conclusion, the authors of Ref. \cite{OST2013} state that a ``complete consistency of the quantum-mechanical and classical descriptions of spinning particles is also established using the Hamiltonian approach in Sec. IV B". However, the quantum-mechanical description --- in particular, the Hermitian Hamiltonian operator, be it $\mathcal{H}$ (2.15) or $\mathcal{H}_\mathrm{FW}$ (3.11), as well as the spin equation of motion (3.15) --- is seriously non-unique as demonstrated at Points \hyperref[H vs H']{({\bf 3})}, \hyperref[FW]{({\bf 4})} and \hyperref[Spin operators]{({\bf 5})} above. Whereas, the classical description of Sect. IV B is unique as we just saw. Thus, the ambiguity of the quantum-mechanical description does not survive in the classical limit. The ambiguity of the covariant Dirac theory regards the energy mean values and eigenvalues \cite{A43,A50}, but the probability current and its motion are unambiguous. In the wave packet approximation, the latter motion can be rewritten as the exact equations of motion of classical (non-spinning) particles in the electromagnetic field, without any use of the semi-classical limit $\hbar\rightarrow 0$ \cite{A46}.

\section{Remarks on the relevant energy concept}\label{Energy}

Previous work emphasized the non-uniqueness of the Dirac {\it energy operator} $\mathrm{E}$, i.e. the Hermitian part of the Hamiltonian operator: it was noted the non-uniqueness of the spectrum of $\mathrm{E}$, as well as that of its general mean values \cite{A43, A50}. At Point \hyperref[H vs H']{({\bf 3})} in the foregoing section, that non-uniqueness was confirmed by an explicit calculation in the most general case. Here we want to indicate why, in this most general situation, the energy operator $\mathrm{E}$ and its mean values {\it ought to be} well defined in a given reference frame. This needs a brief discussion of the relevant concept of energy.

\paragraph{1. Energy in classical physics: continuum vs. test particle. ---}\label{Classical energy} Consider first classical continuum physics and, to begin with, Newtonian continuum physics. There, we have undoubtedly a consistent concept of energy, including a local conservation equation for the total energy density $w$ and its flux $\Mat{\Phi }$. It has the same form as the continuity equation which expresses the conservation of mass:
\be\label{LocalEnergyConservation}
\frac{\partial w}{\partial t}+\mathrm{div}\,\Mat{\Phi }=0.
\ee
This local conservation equation can be integrated to give a global conserved energy $E=\int w \,\dd ^3{\bf x}=\mathrm{Constant}$ --- although this global step needs assuming that the distribution of matter is spatially bounded so that $w$ and $\Mat{\Phi }$ tend to zero quickly enough at infinity. One can take Newtonian gravity as an explicit example \cite{A15}. Now consider a {\it test particle} in Newtonian physics: a piece of matter which is modelled as a point and which is assumed to be passive, i.e., its contribution to the force fields is neglected. Take the case that in an inertial reference frame $\mathrm{R}$, the force over the test particle derives from a potential $V({\bf x},t)$: 
\be\label{F-potential}
{\bf F}\equiv m\frac{\dd {\bf v}}{\dd t}=-\frac{\partial V}{\partial {\bf x}},
\ee
with $m$ the mass of the test particle, ${\bf x}={\bf x}(t)$ its spatial position in the inertial reference frame $\mathrm{R}$, and ${\bf v}=\dd {\bf x}/\dd t$ the velocity of the test particle in $\mathrm{R}$. The total energy $e $ of the test particle is the sum of its kinetic energy and its potential energy $V$ in the external force field:
\be\label{Energy-TestParticle}
e =\frac{1}{2}m{\bf v}^2+V\equiv T+V.
\ee
In the general case, $V=V({\bf x},t)$ depends on the spatial position ${\bf x}$ in $\mathrm{R}$ and on the time $t$. Then, the energy $e $ is of course {\it not} conserved, for we have from (\ref{F-potential}):
\be
\frac{\dd}{\dd t}\left(\frac{1}{2}m{\bf v}^2\right)=m{\bf v.}\frac{\dd {\bf v}}{\dd t} = -\frac{\partial V}{\partial {\bf x}}{\bf .}\frac{\dd {\bf x}}{\dd t}=\frac{\partial V}{\partial t}-\frac{\dd V}{\dd t},
\ee
whence from (\ref{Energy-TestParticle}):
\be
\frac{\dd e}{\dd t}=\frac{\partial V}{\partial t}.
\ee
However, keeping in mind the continuum formulation with conserved energy, $e $ has the clear physical meaning of being the energy of some definite piece of matter that (for some problems) is small enough that it can be considered as a test particle. Note that $e $ is also the value of the Hamiltonian function of the test particle:
\be\label{Hamilton-TestParticle}
e =\frac{{\bf p}^2}{2m}+V({\bf x},t)=H({\bf p},{\bf x},t),\qquad {\bf p}\equiv m{\bf v},
\ee
with which the motion of the test particle can be rewritten as Hamilton's equations. More generally, the total energy of a system of interacting particles in a variable external field is the value of the relevant Hamiltonian function, and it depends on time.\\

\noi In the same way, in special-relativistic continuum physics, the total energy density and its flux can be defined and obey a local equation of conservation \cite{L&L}. Now, be it in a flat or a curved spacetime, the relativistic equation of motion of a test particle in any given coordinate system can be written as Hamilton's equations with a Hamiltonian function of the same standard kind as in non-relativistic physics, i.e., a function $H({\bf p},{\bf x},t)$ with ${\bf p}$ and ${\bf x}$ three-dimensional vectors defined in the given coordinate system. This is well known for a test particle moving in an electromagnetic field in a Minkowski spacetime \cite{Johns2005} and is known also for a geodesic particle in an arbitrary Lorentzian spacetime \cite{Bertschinger1999}. It is still true for a test particle moving in an electromagnetic field in an arbitrary Lorentzian spacetime \cite{A46}. Namely, in the latter general situation, the energy $ e $ is given by $e \equiv c \breve{p}_0$, where $\breve{p}_\mu \equiv g_{\mu \nu} \breve{p}^\nu $ [signature $(+---)$], with \{Eq. (36) in Ref. \cite{A46}\}:
\be \label{Def p^mu}
\breve{p}^{\mu } \equiv mc u^{\mu }   +  \frac{q}{c}  V^{\mu }
\ee
and $u^\mu \equiv \dd x^\mu /\dd s$ with $s=c\tau $ where $\tau $ is the proper time along the trajectory of the particle, $V^{\mu }$ being the electromagnetic potential and $q$ the charge of the test particle. This same $e\equiv c \breve{p}_0 $ can be extracted from the energy-momentum relation
\be\label{44 Paper 4}
 g^{\mu \nu }  \left( \breve{p}_{\mu }   -  \frac{q}{c}  V_{\mu }  \right) \left( \breve{p}_{\nu }   -  \frac{q}{c}  V_{\nu }  \right) - m^{2} c^{2} =0
\ee
as a function 
\be\label{def H test particle}
H\left( {\bf p} , {\bf x}, t \right) \equiv e \equiv  c\breve{p}_0
\ee
with ${\bf p}=(p_j)\equiv (-\breve{p}_j)$. This is indeed a Hamiltonian function for the motion of the test particle, as was briefly shown in Ref. \cite{A46}, Eqs. (79)--(80). The Hamiltonian function of a geodesic particle \cite{Bertschinger1999} is the case $V_\mu =0$. \\

\paragraph{2. The classical energy depends on the reference frame. ---}\label{F-dependence-classical} 

The dependence of the energy of a classical test particle on the reference frame is obvious and well known in the case of a free particle in Newtonian mechanics [$V=0$ in Eqs. (\ref{F-potential})--(\ref{Hamilton-TestParticle})]: such a particle has only kinetic energy, and that energy of course depends on the reference frame. Two inertial reference frames $\mathrm{R}$ and $\mathrm{R}'$ are defined by the spatial position charts $X\mapsto {\bf x}\equiv (x^j)$ and $X\mapsto {\bf x}'\equiv (x'^j)$ that each among $\mathrm{R}$ and $\mathrm{R}'$ assigns to any event $X$ in the Galilean spacetime. Let ${\bf V}$ be the constant velocity with which $\mathrm{R}'$ is moving with respect to $\mathrm{R}$, i.e., ${\bf x}={\bf x'}+ {\bf V}t$. The velocities of a particle in $\mathrm{R}$ and $\mathrm{R}'$, ${\bf v}\equiv \frac{\dd {\bf x}}{\dd t}$ and ${\bf v}'\equiv \frac{\dd {\bf x}'}{\dd t}$, are related by ${\bf v}={\bf v}'+{\bf V}$. The difference in the respective (kinetic) energies is $e-e'=\frac{1}{2}m{\bf V}^2+m{\bf v}'{\bf .V}$. 
\footnote{\ 
It is easy to recover that difference from the relativistic expression (\ref{def H test particle}), in the non-relativistic limit.
}
This difference can be felt very concretely. Next, consider the case $V\ne0$. When going from $\mathrm{R}$ to another inertial frame $\mathrm{R}'$, the potential force field is left unchanged, in the sense that $V'({\bf x}',t)=V({\bf x},t)$ for ${\bf x}={\bf x'}+ {\bf V}t$. Hence, here again, the difference in the energies $e$ and $e'$ comes from the difference in the kinetic energies $T$ and $T'$, which is very concrete.\\

Also in a general relativistic spacetime, the energy of a test particle depends on the general coordinate system, because $e/c$ in Eq. (\ref{def H test particle}) is the time component of the spacetime covector $\breve{p}_\mu $. However, for the same reason, $e $ stays invariant under a purely spatial coordinate change (\ref{purely-spatial-change}). Thus, {\it the energy of a test particle in a general spacetime depends precisely on the reference frame} as defined in Section \ref{Hamiltonians}, i.e., on the equivalence class of charts defined on a given open domain U of the spacetime manifold V, modulo the changes (\ref{purely-spatial-change}). \\

We conclude that the definition of the energy of a classical test particle as being the value of its Hamiltonian function: $e \equiv H({\bf p},{\bf x}, t)$ with $H$ being given by Eq. (\ref{Hamilton-TestParticle}) in Newtonian physics and by (\ref{def H test particle}) in relativistic physics, is unique in any given reference frame. This energy is not conserved in general. For relativistic physics, these conclusions are valid in a general reference frame in a general spacetime, and also in the presence of an electromagnetic field. The uniqueness of $e $ must be understood for a given field of the external potential ($V$ or $V_\mu $), when there is one. (There is none for free motion in a Galilean spacetime, nor for geodesic motion in a curved spacetime.) A very important point recalled above is that the energy of a classical test particle does depend on the reference frame. We consider this dependence as physically clear. We do not see it as a non-uniqueness {\it problem} in any sensible meaning. 

\paragraph{3. Energy operator in a general spacetime. ---}\label{QM-Energy} 

It is well known that QM has a strong relation to classical Hamiltonian mechanics. This is already apparent in the fact that the non-relativistic Schr\"odinger equation for a single particle is got from the non-relativistic classical Hamiltonian (\ref{Hamilton-TestParticle}) of a test particle, by applying the classical-quantum correspondence
\be\label{Classical-quantum}
e  \rightarrow  +\iC\hbar \frac{\partial }{\partial t}, \qquad p_j \rightarrow -\iC\hbar \frac{\partial }{\partial x^j}.
\ee
The wave equations of relativistic QM can also be obtained by this same method, e.g., as is well known, the Klein-Gordon equation. 
The correspondence (\ref{Classical-quantum})$_1$ leads also directly to interpret the Hamiltonian operator $\mathrm{H}$ in the general Schr\"odinger equation (\ref{Schrodinger-general}) as the {\it energy operator,} whose eigenvalues, which are associated with stationary solutions of (\ref{Schrodinger-general}), are the observable values of the energy of the quantum-mechanical system at hand. The meaning of the energy of that system is hence inherited from classical Hamiltonian mechanics. 
In particular, in the case of a one-particle system, the energy operator is the quantum-mechanical equivalent of the energy of a test particle. However, in the case of the covariant Dirac equation, it happens that the operator $\mathrm{H}$ in general is not Hermitian for the relevant scalar product \cite{A43,Parker1980,Leclerc2006}. Hence, for the covariant Dirac equation one defines the energy operator $\mathrm{E}$ precisely as the Hermitian part of $\mathrm{H}$ for the relevant scalar product \cite{A43,Leclerc2006}; $\mathrm{E}$ has another important property which has usually $\mathrm{H}$ (when $\mathrm{H}$ is Hermitian): its mean value is the field energy, i.e. the space integral of the $T^0_0$ component of the (canonical) energy-momentum tensor, in that case that of the ``classical" Dirac field \cite{A43,A48,Leclerc2006}. \\

We saw in Sect. \ref{Hamiltonians} that, in a general spacetime, the Hamiltonian operator $\mathrm{H}$ (for a fixed choice of the gauge fields $G$) depends on the reference frame as we define it --- just as does the classical Hamiltonian (\ref{def H test particle}) of a relativistic test particle, as we saw at \hyperref[F-dependence-classical]{Point 2 above}. 
It follows from the definition of the energy operator $\mathrm{E}$ as the Hermitian part of $\mathrm{H}$ that this dependence on the reference frame applies also to $\mathrm{E}$. In a variable external field, the energy of a classical test particle, or the total energy of a system of interacting particles, depend on time. In the same way, the mean values of the energy operator in a given reference frame $\mathrm{F}$ do depend on the coordinate time $t=x^0/c$ if the gravitational field is not stationary in $\mathrm{F}$ --- i.e., if the metric components $g_{\mu \nu}$ depend on $x^0$, in charts belonging to $\mathrm{F}$. In such a general case, there is no stationary state in the standard sense, but still, at each given time $t_0$ one may define an energy spectrum and a set $\mathcal{S}_{t_0}$ of ``stationary states" by considering the ``frozen" energy operator $\mathrm{E}_{t_0}$ of that time \cite{A43}. Recall that according to QM: if a particle is in a state $\Psi $ at time $t_0$, then an energy measurement at that time will yield an eigenvalue of $\mathrm{E}_{t_0}$, arising together with the collapse of the state $\Psi $ into an eigenvector of $\mathrm{E}_{t_0}$. \\

\paragraph{4. Dependence on reference frame vs. dependence on tetrad field. ---}\label{Dirac-E-tetrad} 

The data of a {\it reference frame} F, in the specific sense \hyperref[Def-Ref-Frame]{meant here}, provides all information about the relevant congruence of ``observers" (or rather: the congruence of spatial reference points, each of them being given by a time-like world line having constant spatial coordinates in any chart of F), and about the relevant time function $t$. This data contains indeed any information that is needed to specify the motion of the apparatus/laboratory --- each point of which being assumed either to be fixed (i.e. having fixed spatial coordinates $x^j$) in F, or to have a well defined motion (i.e. a well-defined dependence $x^j=x^j(t)$). In particular, the data of a reference frame defines the rotation-rate field of the reference points \cite{A47,Weyssenhoff1937,Cattaneo1958} --- for instance, that field may be a uniform rotation (or a non-uniform one as well) with respect to an inertial frame in a Minkowski spacetime. The choice of the reference frame is mainly fixed by the experimental/observational settings, essentially in the same way in classical mechanics and in quantum mechanics. As a concrete example, we may consider very slow neutrons in a laboratory at the surface of the rotating Earth (its rotation velocity being assumed uniform and constant, thus the metric in the rotating frame is stationary) and assume that some device puts them in stationary energy states in that laboratory, like in the experiments of Nesvizhevsky {\it et al.} \cite{Nesvizhevsky2002}. Then, a uniformly rotating frame is relevant for discussing the effect of the Earth's rotation on the energy levels \cite{A41}. (Note that  for a wave function, the form: $\Psi (t,{\bf x}) = \phi (t)\,A({\bf x})$, which is a prerequisite for $\Psi $ to be a stationary state, depends again exactly on the reference frame.) As another example, consider some spacecraft. If a neutron interferometer or some other device able to do quantum measurements is on board, then this spacecraft itself defines at least a very local reference frame, which is the relevant one for the measurements made on board. Of course, in order to have precise data about timing and about its position in space, the spacecraft will receive informations from distant observers not rigidly bound to its local reference frame. However, basically the relevant energy operator is the one in a reference frame that locally coincides with the spacecraft (but that should possibly have to be extended away from it).\\

The additional data of a {\it tetrad field} $(u_\alpha )$ includes that of the 4-velocity field $u_0$. This vector field determines another congruence of world lines (the integral curves of $u_0$), which in general is {\it different} from the congruence specified by the given reference frame. The data of a tetrad field involves also the data of the spatial triad field $(u_p)_{p=1,2,3}$, which may have an arbitrary and {\it independent} rotation with respect to the congruence of world lines determined by the vector field $u_0$ of the tetrad (and {\it a fortiori} with respect to the congruence specified by the given reference frame). See Sect. 3 in Ref. \cite{A49}. Thus the data of a tetrad field seems entirely superfluous, e.g. to determine the stationary states of neutrons in an Earth-based laboratory (the situation outlined above).\\ 

Therefore, in our opinion, the dependence of the energy operator $\mathrm{E}$ on the tetrad field \cite{A43}, which has been shown explicitly in the general case in Section \ref{OST} above, and which appears in particular in the calculation of the Mashhoon effect for a Dirac particle \cite{A49}, is not a physical effect. Instead, we consider that dependence as an undesirable effect of the gauge freedom of the covariant Dirac equation. Thus, the choice of the tetrad field should be restricted in such a way that, in a given reference frame, all remaining choices lead to equivalent energy operators. Two different ways to do this have been proposed \cite{A47,A48}.

\paragraph{5. Summary for the energy operator. ---}\label{Summary E} The energy operator $\mathrm{E}$ of a one-particle system corresponds with the energy $e$ of a test particle by the classical-quantum correspondence (\ref{Classical-quantum}) and, just as does $e$, it depends on the reference frame and evolves with the coordinate time. It acts on functions depending on the spatial position, which is restricted to the spatial range $\Omega $ of a coordinate system belonging to a given reference frame. [It would be better to regard the spatial position as an element (point) of the space manifold associated with that reference frame; see Note \ref{SpaceManifold}.] This restriction is not a problem, because the domain of the coordinate system can be taken large enough that the wave function can safely be assumed to vanish at the boundary of $\Omega $. The present arguments show that the energy operator $\mathrm{E}$ ought to be well defined in a given reference frame at a given time. However, as we saw in Section \ref{Hamiltonians}, this is not the case if the wave equation depends on a gauge choice leading to time-dependent unitary transformations. Unfortunately, the latter thing is what happens with the covariant Dirac equation, even in the Schwinger gauge, as was confirmed by an explicit study in the most general situation at Section \ref{OST}.\\

\noi {\bf Acknowledgement.} A referee asked questions about the difference between classical and quantum mechanics for what regards the non-uniqueness of the Hamiltonian, and about the meaning of that non-uniqueness for quantum mechanics, both theoretically and experimentally. This led me to  emphasize or to clarify several points regarding reference frames, the definition of the classical energy and its dependence on the reference frame, and the meaning of the dependence of the Dirac energy operator on the tetrad field.


\begin{thebibliography}{9}
\small

\bibitem{BrillWheeler1957}
D. R. Brill and J. A. Wheeler, Rev. Modern Phys. {\bf 29}, 465--479 (1957). 

\bibitem{BrillWheeler1961}
D. R. Brill and J. A. Wheeler, Erratum: Rev. Modern Phys. {\bf 33}, 623--624 (1961).

\bibitem{ChapmanLeiter1976}
T. C. Chapman and D. J. Leiter, Am. J. Phys. {\bf 44}, No. 9, 858--862 (1976).

\bibitem{Fock1929}
V. Fock, J. Phys. Rad. {\bf 10}, No. 11, 392--405 (1929).

\bibitem{A43}
M. Arminjon and F. Reif\mbox{}ler, Ann. Phys. (Berlin) {\bf 523}, 531--551 (2011). 

\bibitem{A50}
M. Arminjon, Int. J. Theor. Phys. {\bf 52}, 4032--4044 (2013). 

\bibitem{Ryder2008}
L. Ryder, Gen. Relativ. Gravit. {\bf 40}, 1111--1115 (2008).

\bibitem{A49} 
M. Arminjon, Int. J. Theor. Phys. {\bf 53}, 1993--2013 (2014). 

\bibitem{GorbatenkoNeznamov2011}
M. V. Gorbatenko and V. P. Neznamov, Phys. Rev. D {\bf 83}, 105002 (2011). 

\bibitem{GorbatenkoNeznamov2013}
M. V. Gorbatenko and V. P. Neznamov, preprint arXiv:1301.7599v2 (gr-qc). 

\bibitem{OST2013} 
Yu. N. Obukhov, A. J. Silenko, and O. V. Teryaev, Phys. Rev. D {\bf 88}, 084014 (2013). 

\bibitem{A47}
M. Arminjon, Ann. Phys. (Berlin) {\bf 523}, 1008--1028 (2011). 

\bibitem{A48}
M. Arminjon, Int. J. Geom. Meth. Mod. Phys. {\bf 10}, No. 7, 1350027 (2013). 

\bibitem{GorbatenkoNeznamov2011b}
M. V. Gorbatenko and V. P. Neznamov, preprint arXiv:1107.0844v6 (gr-qc). 

\bibitem{GorbatenkoNeznamov2014}
M. V. Gorbatenko and V. P. Neznamov, Ann. Phys. (Berlin) {\bf 526}, 195--200 (2014).

\bibitem{A46}
M. Arminjon and F. Reif\mbox{}ler, Braz. J. Phys. {\bf 43},  64--77 (2013). 

\bibitem{A44}
M. Arminjon and F. Reif\mbox{}ler, Int. J. Geom. Meth. Mod. Phys. {\bf 8}, 155--165 (2011). 

\bibitem{FoldyWouthuysen1950}
L. L. Foldy and S. A. Wouthuysen, Phys. Rev. {\bf 78}, 29--36 (1950).

\bibitem{Goldman1977}
T. Goldman, Phys. Rev. D {\bf 15}, 1063--1067 (1977).


\bibitem{Eriksen1958}
E. Eriksen, Phys. Rev. {\bf 111}, 1011--1016 (1958).

\bibitem{Schulten2000}
K. Schulten, {\it Notes on quantum mechanics,} \href{http://www.ks.uiuc.edu/Services/Class/PHYS480/qm_PDF/QM_Book.pdf}{online course} (2000), Eq. (10.311).

\bibitem{A15} M. Arminjon, Arch. Mech. {\bf 48}, No. 1, 25--52 (1996).

\bibitem{L&L} L. Landau and E. Lifchitz, {\it Th\'eorie des Champs} (Fourth French edn., Mir, Moscow 1989), Sections 31--33. (Fourth English edition: E. M. Lifshitz and L. D. Landau, {\it The Classical Theory of Fields}, Butterworth Heinemann, Oxford 1980.)

\bibitem{Johns2005}
O. D. Johns, \textit{Analytic Mechanics for Relativity and Quantum Mechanics} (Oxford University Press, 2005).

\bibitem{Bertschinger1999}
E. Bertschinger,  {\it Hamiltonian dynamics of particle motion,} \href{http://web.mit.edu/edbert/GR/gr3.pdf}{online course} (1999).

\bibitem{L&L Mechanics}
L. Landau and E. Lifchitz, {\it M\'ecanique} (Fourth French edn., Mir, Moscow 1982), Section 45. (Third English edition: L. D. Landau and E. M. Lifshitz, {\it Mechanics}, Butterworth Heinemann, Oxford 1976.)


\bibitem{Parker1980}
L. Parker, Phys. Rev. D {\bf 22}, 1922--1934 (1980).

\bibitem{Leclerc2006}
M. Leclerc, Class. Quant. Grav. {\bf 23}, 4013--4020 (2006). 

\bibitem{Weyssenhoff1937}
J. von Weyssenhof, Bull. Acad. Polon. Sci., Sect. A, 252 (1937). (Quoted by Cattaneo \cite{Cattaneo1958}.)

\bibitem{Cattaneo1958}
C. Cattaneo, Nuovo Cim. {\bf 10}, 318--337 (1958).



\bibitem{Nesvizhevsky2002} 
V. V. Nesvizhevsky {\it et al.,} Nature {\bf 415}, 297--299 (2002).

\bibitem{A41}
M. Arminjon, Phys. Lett. A {\bf 372}, 2196--2200 (2008). 







\end{thebibliography}
\end{document}